\documentclass[11pt,a4paper,]{article}
\usepackage{lmodern}

\usepackage{amssymb,amsmath}
\usepackage{ifxetex,ifluatex}
\usepackage{fixltx2e} 
\ifnum 0\ifxetex 1\fi\ifluatex 1\fi=0 
  \usepackage[T1]{fontenc}
  \usepackage[utf8]{inputenc}
\else 
  \usepackage{unicode-math}
  \defaultfontfeatures{Ligatures=TeX,Scale=MatchLowercase}
\fi
\IfFileExists{upquote.sty}{\usepackage{upquote}}{}
\IfFileExists{microtype.sty}{%
\usepackage[]{microtype}
\UseMicrotypeSet[protrusion]{basicmath} 
}{}
\PassOptionsToPackage{hyphens}{url} 
\usepackage[unicode=true]{hyperref}
\hypersetup{
            pdftitle={A new tidy data structure to support exploration and modeling of temporal data},
            pdfkeywords={time series, data wrangling, tidy data, R, forecasting, data science, exploratory data analysis, data pipelines},
            pdfborder={0 0 0},
            breaklinks=true}
\usepackage{geometry}
\usepackage[style=authoryear-comp,]{biblatex}
\usepackage{longtable,booktabs}
\IfFileExists{footnote.sty}{\usepackage{footnote}\makesavenoteenv{long table}}{}
\IfFileExists{parskip.sty}{%
\usepackage{parskip}
}{
\setlength{\parindent}{0pt}
\setlength{\parskip}{6pt plus 2pt minus 1pt}
}
\providecommand{\tightlist}{%
  \setlength{\itemsep}{0pt}\setlength{\parskip}{0pt}}
\makeatletter
\def\fps@figure{htbp}
\makeatother

\title{A new tidy data structure to support exploration and modeling of temporal data}

\RequirePackage{caption}
\DeclareCaptionStyle{italic}[justification=centering]
 {labelfont={bf},textfont={it},labelsep=colon}
\RequirePackage{bera}
\RequirePackage{mathpazo}

\RequirePackage{fancyhdr}
\fancypagestyle{plain}{%
\fancyhf{} 
\fancyfoot[C]{\sffamily\thepage} 

}

\RequirePackage{bm,amsmath}
\allowdisplaybreaks

\RequirePackage{graphicx}
\RequirePackage[compact,sf,bf]{titlesec}
\titleformat{\section}[block]
  {\fontsize{15}{17}\bfseries\sffamily}
  {\thesection}
  {0.4em}{}
\titleformat{\subsection}[block]
  {\fontsize{12}{14}\bfseries\sffamily}
  {\thesubsection}
  {0.4em}{}
\titlespacing{\section}{0pt}{*5}{*1}
\titlespacing{\section}{0pt}{*2}{*0.2}

\def\Date{\number\day}
\def\Month{\ifcase\month\or
 January\or February\or March\or April\or May\or June\or
 July\or August\or September\or October\or November\or December\fi}
\def\Year{\number\year}

\makeatletter
\def\wp#1{\gdef\@wp{#1}}\def\@wp{??/??}
\def\jel#1{\gdef\@jel{#1}}\def\@jel{??}
\def\showjel{{\large\textsf{\textbf{JEL classification:}}~\@jel}}
\def\nojel{\def\showjel{}}
\def\addresses#1{\gdef\@addresses{#1}}\def\@addresses{??}
\def\cover{{\sffamily\setcounter{page}{0}
        \thispagestyle{empty}%
        \vspace*{-2cm}
        \centerline{\raisebox{-1.8cm}{\includegraphics[width=5cm]{MBSportrait}}\hspace*{9cm} ISSN 1440-771X}\vspace{0.99cm}
        \begin{center}\Large
        Department of Econometrics and Business Statistics\\[.5cm]
        \scriptsize http://business.monash.edu/econometrics-and-business-statistics/research/publications
        \end{center}\vspace{2cm}
        \begin{center}
        \fbox{\parbox{14cm}{\begin{onehalfspace}\centering\Huge\vspace*{0.3cm}
                \textsf{\textbf{\expandafter{\@title}}}\vspace{1cm}\par
                \LARGE\@author\end{onehalfspace}
        }}
        \end{center}
        \vfill
                \begin{center}\Large
                \Month~\Year\\[1cm]
                Working Paper \@wp
        \end{center}}}
\def\pageone{{\sffamily
        \newpage
        \thispagestyle{empty}
        \vbox to 23cm{
        \raggedright\baselineskip=1.2cm
     {\fontsize{24.88}{30}\sffamily\textbf{\expandafter{\@title}}}
        \vspace{2cm}\par
        \hspace{1cm}\parbox{14cm}{\sffamily\large\@addresses}\vspace{1cm}\vfill
        \hspace{1cm}{\large\Date~\Month~\Year}\\[1cm]
        \hspace{1cm}\showjel\vss}}}
\def\blindtitle{{\sffamily
     \thispagestyle{plain}\raggedright\baselineskip=1.2cm
     {\fontsize{24.88}{30}\sffamily\textbf{\expandafter{\@title}}}\vspace{1cm}\par
        }}
\def\titlepage{{\cover\newpage\setstretch{1}\pageone\newpage\blindtitle}}

\def\blind{\def\titlepage{{\blindtitle}}\let\maketitle\blindtitle}
\def\titlepageonly{\def\titlepage{{\pageone\end{document}}}}
\def\nocover{\def\titlepage{{\newpage\setstretch{1}\pageone\newpage\blindtitle}}\let\maketitle\titlepage}
\let\maketitle\titlepage
\makeatother

\RequirePackage{setspace}
\spacing{1.5}

\RequirePackage{microtype}

\RequirePackage{xcolor} 
\definecolor{darkblue}{rgb}{0,0,.6}
\RequirePackage{url}

\makeatletter
\@ifpackageloaded{hyperref}{}{\RequirePackage{hyperref}}
\makeatother
\hypersetup{
     citecolor=0 0 0,
     breaklinks=true,
     bookmarksopen=true,
     bookmarksnumbered=true,
     linkcolor=darkblue,
     urlcolor=blue,
     citecolor=darkblue,
     colorlinks=true}

\newenvironment{keywords}{\par\vspace{0.5cm}\noindent{\sffamily\textbf{Keywords:}}}{\vspace{0.25cm}\par\hrule\vspace{0.5cm}\par}

\renewenvironment{abstract}{\begin{minipage}{\textwidth}\parskip=1.4ex\noindent
\hrule\vspace{0.1cm}\par{\sffamily\textbf{\abstractname}}\newline}
  {\end{minipage}}

\usepackage[T1]{fontenc}
\usepackage[utf8]{inputenc}

\usepackage[showonlyrefs]{mathtools}
\usepackage[no-weekday]{eukdate}

\makeatletter
\@ifpackageloaded{biblatex}{}{\usepackage[style=authoryear-comp, backend=biber, natbib=true]{biblatex}}
\makeatother
\DeclareFieldFormat{url}{\texttt{\url{#1}}}
\DeclareFieldFormat[article]{pages}{#1}
\DeclareFieldFormat[inproceedings]{pages}{\lowercase{pp.}#1}
\DeclareFieldFormat[incollection]{pages}{\lowercase{pp.}#1}
\DeclareFieldFormat[article]{volume}{\mkbibbold{#1}}
\DeclareFieldFormat[article]{number}{\mkbibparens{#1}}
\DeclareFieldFormat[article]{title}{\MakeCapital{#1}}
\DeclareFieldFormat[article]{url}{}
\DeclareFieldFormat[inproceedings]{title}{#1}
\DeclareFieldFormat{shorthandwidth}{#1}
\usepackage{xpatch}
\xpatchbibmacro{volume+number+eid}{\setunit*{\adddot}}{}{}{}
\renewbibmacro{in:}{%
  \ifentrytype{article}{}{%
  \printtext{\bibstring{in}\intitlepunct}}}

\AtEveryBibitem{\clearfield{month}}
\AtEveryCitekey{\clearfield{month}}

\makeatletter
\DeclareDelimFormat[cbx@textcite]{nameyeardelim}{\addspace}
\makeatother

\wp{no/yr}
\nojel

\nocover

\author{Earo~Wang, Dianne~Cook, Rob J~Hyndman}
\addresses{\textbf{Earo Wang}\newline
Department of Econometrics and Business Statistics, \newline Monash University, VIC 3800 \newline Australia.
\newline{Email: \href{mailto:earo.wang@monash.edu}{\nolinkurl{earo.wang@monash.edu}}}\newline Corresponding author\\[1cm]
\textbf{Dianne Cook}\newline
Department of Econometrics and Business Statistics, \newline Monash University, VIC 3800 \newline Australia.
\newline{Email: \href{mailto:dicook@monash.edu}{\nolinkurl{dicook@monash.edu}}}\\[1cm]
\textbf{Rob J Hyndman}\newline
Department of Econometrics and Business Statistics, \newline Monash University, VIC 3800 \newline Australia.
\newline{Email: \href{mailto:rob.hyndman@monash.edu}{\nolinkurl{rob.hyndman@monash.edu}}}\\[1cm]
}

\date{\sf\Date~\Month~\Year}

\usepackage{todonotes}
\usepackage{multirow}
\usepackage{booktabs}
\usepackage{animate}
\usepackage[section]{placeins}

\usepackage[english]{babel}
\addto\extrasenglish{

}

\begin{document}
\maketitle
\begin{abstract}
Mining temporal data for information is often inhibited by a multitude of formats: irregular or multiple time intervals, point events that need aggregating, multiple observational units or repeated measurements on multiple individuals, and heterogeneous data types. On the other hand, the software supporting time series modeling and forecasting, makes strict assumptions on the data to be provided, typically requiring a matrix of numeric data with implicit time indexes. Going from raw data to model-ready data is painful. This work presents a cohesive and conceptual framework for organizing and manipulating temporal data, which in turn flows into visualization, modeling and forecasting routines. Tidy data principles are extended to temporal data by: (1) mapping the semantics of a dataset into its physical layout; (2) including an explicitly declared index variable representing time; (3) incorporating a ``key'' comprising single or multiple variables to uniquely identify units over time. This tidy data representation most naturally supports thinking of operations on the data as building blocks, forming part of a ``data pipeline'' in time-based contexts. A sound data pipeline facilitates a fluent workflow for analyzing temporal data. The infrastructure of tidy temporal data has been implemented in the R package \textbf{tsibble}.
\end{abstract}

\hypertarget{sec:intro}{%
\section{Introduction}\label{sec:intro}}

Temporal data arrives in many possible formats, with many different time contexts. For example, time can have various resolutions (hours, minutes, and seconds), and can be associated with different time zones with possible adjustments such as summer time. Time can be regular (such as quarterly economic data or daily weather data), or irregular (such as patient visits to a doctor's office). Temporal data also often contains rich information: multiple observational units of different time lengths, multiple and heterogeneous measured variables, and multiple grouping factors. Temporal data may comprise the occurrence of events, such as flight departures, that need to be reduced to a regular structure.

Despite this variety and heterogeneity of temporal data, current software typically requires time series objects to be model-oriented matrices. Analysts are expected to do their own data preprocessing and take care of anything else needed to allow model fitting, which leads to a myriad of ad hoc solutions and duplicated efforts.

\begin{figure}

{\centering \includegraphics[width=\textwidth]{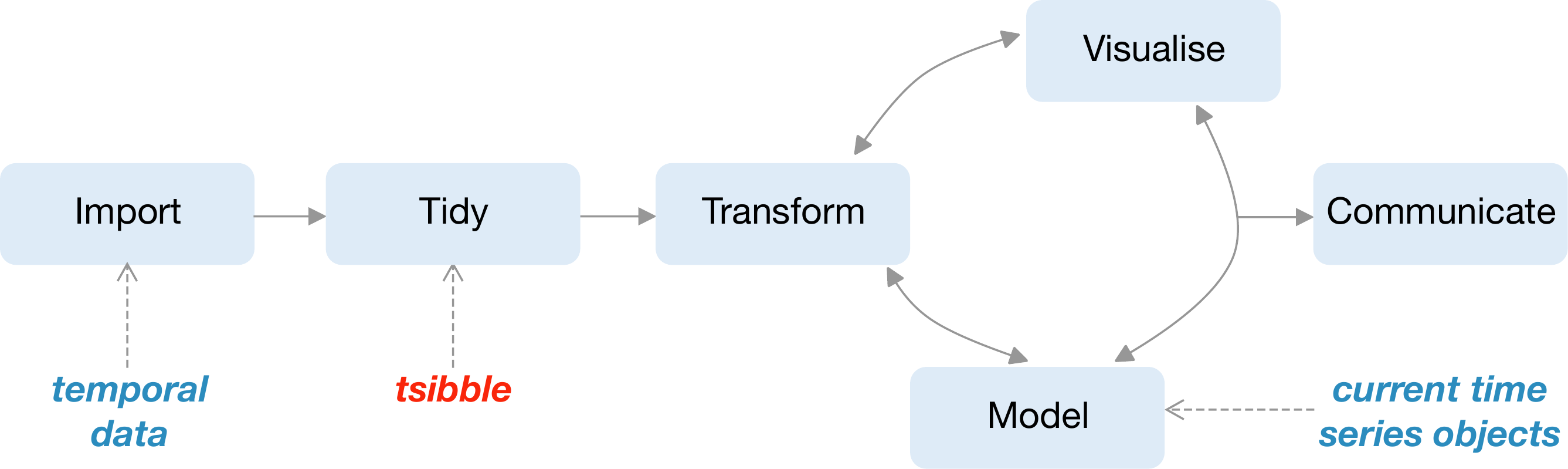} 

}

\caption{Illustration of the data science workflow, drawn from \textcite{r4ds}, showing how current time series tools interface with the workflow and how the tsibble structure and tools integrate. The new data structure, tsibble, makes the connection between temporal data input, and specialist modeling formats. It provides elements at the ``tidy'' step, which produce tidy temporal data for time series visualization and modeling.}\label{fig:tsibble-fit}
\end{figure}

\textcite{r4ds} proposed the tidy data workflow, which provides a conceptual
framework for processing data (as described in \autoref{fig:tsibble-fit}). Currently, time series modeling and forecasting enters this framework at the \emph{modeling} stage, while temporal data enters at the start. This paper integrates
time series analysis into this tidy framework, providing a coherent way for getting temporal data into the matrix format for modeling.

The paper is structured as follows. \autoref{sec:str} reviews temporal data structures corresponding to time series and longitudinal analysis, and discusses ``tidy data'' and the grammar of data manipulation. \autoref{sec:semantics} proposes contextual semantics for temporal data, built on top of tidy data. The concept of data pipelines with respect to the time domain will be discussed in depth in \autoref{sec:pipeline}, followed by a discussion of the design choices made in the software structure in \autoref{sec:software}. Two case studies are presented in \autoref{sec:cases} illustrating temporal data exploration using the newly implemented infrastructure. \autoref{sec:conclusion} discusses future work.

\hypertarget{sec:str}{%
\section{Data structures}\label{sec:str}}

\hypertarget{sec:time-series}{%
\subsection{Time series and longitudinal data}\label{sec:time-series}}

Temporal data problems are typically grouped into two types of analysis, time series and longitudinal. Despite being exactly the same data input, the representation of time series and longitudinal data diverges due to different modeling approaches.

Time series can be univariate or multivariate, and for modeling require relatively long lengths (i.e., large \(T\)). Time series researchers and analysts who are concerned with this large \(T\) property, are mostly concerned with stochastic processes, for the primary purpose of forecasting, and characterizing temporal dynamics. Most statistical software represent such time series as vectors or matrices. Multivariate time series are typically assumed to be in the format where each row is assumed to hold observations at a time point and each column to contain a single time series. (The tidy data name for this would be \textbf{wide format}.) This implies that data are columns of homogeneous types: numeric or non-numeric, but there are limited supporting methods for non-numeric variables. In addition, time indexes are stripped off the data and implicitly inferred as attributes or meta-information. There is a strict requirement that the number of observations must be the same across all the series. Data wrangling, from the form that data arrives in, to this specialist format, can be frustrating and difficult, inhibiting the performance of downstream tasks.

For longitudinal analysis, researchers and analysts are primarily interested in explaining trends across and variations among individuals, and making inference about a broader population. Longitudinal data or panel data typically assumes fewer measurements (small \(T\)) over a large number of individuals (large \(N\)). It often occurs that measurements for individuals are taken at different time points, resulting in an unbalanced panel. Thus, the primary format required for modeling such data is stacked series, blocks of measurements for each individual, with columns indicating individual, times of measurement and the measurements themselves. (The tidy data name for this would be \textbf{long format}.) Evidently, this data organization saves storage space for many sparse cells, compared to structuring it into wide format which would have missing values in many cells. A drawback of this format is that information unique to each individual is often repeated for all time points. An appealing feature is that data is structured in a semantic manner with reference to observations and variables, with the time index stated explicitly. This opens the door to easily operating on time to make calculations and extract different temporal components, such as month and day of the week. It is conducive to examining the data in many different ways and leading to more comprehensive exploration and forecasting.

\hypertarget{tidy-data-and-the-grammar-of-data-manipulation}{%
\subsection{Tidy data and the grammar of data manipulation}\label{tidy-data-and-the-grammar-of-data-manipulation}}

\textcite{wickham2014tidy} coined the term ``tidy data'', which is a rephrasing of the second and third normal forms in relational databases but in a way that makes more sense to data scientists by referring rows to observations and columns to variables. The principles of ``tidy data'' attempt to standardize the mapping of the semantics of a dataset to its physical representation. This data structure is the fundamental unit of the \textbf{tidyverse}, which is a collection of R packages designed for data science. The ubiquitous use of the \textbf{tidyverse} is testament to the simplicity, practicality and general applicability of the tools. The \textbf{tidyverse} provides abstract yet functional grammars to manipulate and visualize data in easier-to-comprehend form. One of the \textbf{tidyverse} packages, \textbf{dplyr} \autocite{R-dplyr}, showcases the value of a grammar as a principled vehicle to transform data for a wide range of data challenges, providing a consistent set of verbs: \texttt{mutate()}, \texttt{select()}, \texttt{filter()}, \texttt{summarize()}, and \texttt{arrange()}. Each verb focuses on a singular task. Most common data tasks can be rephrased and tackled with these five key verbs, by composing them sequentially.

The \textbf{tidyverse} largely formalizes exploratory data analysis. Many in the R community have adopted the \textbf{tidyverse} way of thinking and extended it to broader domains, such as simple features for spatial data in the \textbf{sf} package \autocite{RJ-2018-009} and missing value handling in the \textbf{naniar} package \autocite{tierney-naniar-2018}. Temporal data tools need to catch up.

\hypertarget{existing-time-series-standards-in-r}{%
\subsection{Existing time series standards in R}\label{existing-time-series-standards-in-r}}

Current standards, provided by the native \texttt{ts} object in R, and extended by \textbf{zoo} \autocite{zoo2005} and \textbf{xts} \autocite{R-xts}, assemble temporal data into matrices with implicit time indexes. These objects were designed for modeling methods. The diagram in the style of \autoref{fig:tsibble-fit} would place the model at the center of the analytical universe, and all the transformations and visualizations would hinge on that format. This is contrary to the \textbf{tidyverse} conceptualization, which holistically captures the full data workflow.

A new temporal data class is needed in the upstream of the workflow, which could incorporate all the downstream modules. A relatively new R package \textbf{tibbletime} \autocite{R-tibbletime} proposed a data class of \emph{time tibble} to represent temporal data in heterogeneous tabular format. It only requires an index variable to declare a temporal data object, thus placing it at the import stage. However, as proposed in \autoref{sec:semantics} a more rigid data structure is required for time series analytics and models.

This paper describes a new tidy representation for temporal data, and a unified framework to streamline the workflow from data preprocessing to visualization and forecasting, as an integral part of a tidy data analysis.

\hypertarget{sec:semantics}{%
\section{Contextual semantics}\label{sec:semantics}}

\begin{figure}

{\centering \includegraphics[width=\textwidth]{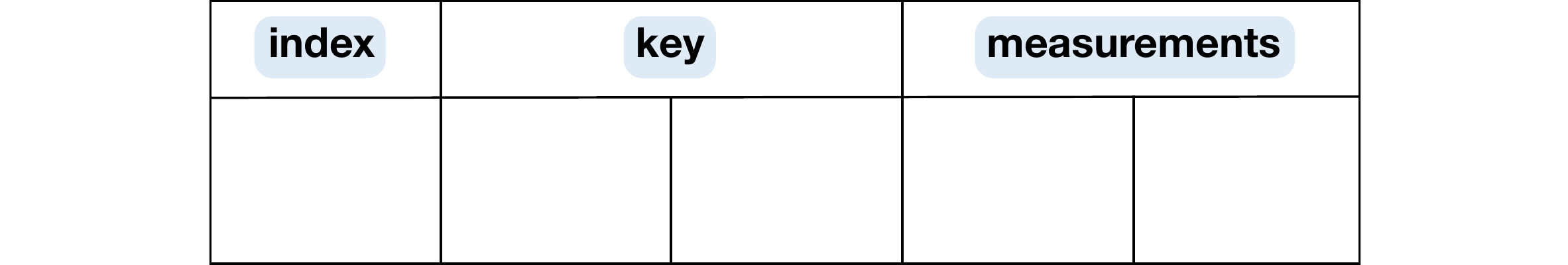} 

}

\caption{The architecture of the tsibble structure is built on top of the data frame---a two-dimensional array in R, with time series semantics: index and key.}\label{fig:tsibble-semantics}
\end{figure}

The choice of tidy representation of temporal data arises from a data-centric perspective, which accommodates all of the operations that are to be performed on the data. \autoref{fig:tsibble-fit} marks where this new abstraction is placed in the tidy model, which we refer to as a ``tsibble''. The tsibble structure is an extension of a data frame---a two-dimensional array in R---with additional time series semantics: index and key, as shown in \autoref{fig:tsibble-semantics}.

To demonstrate the concept of the tsibble, \autoref{tab:tb-sub} presents a subset of tuberculosis cases estimated by \textcite{tb-data}. It contains 12 observations and 5 variables arranged in a ``long'' tabular form. Each observation comprises the number of people who are diagnosed with tuberculosis for each gender at three selected countries in the years of 2011 and 2012. To turn this data into a tsibble: (1) column \texttt{year} is declared as the index variable; (2) the key is specified to consist of columns \texttt{country} and \texttt{gender}. The column \texttt{count} is the only measured variable in this data, but the structure is sufficiently flexible to hold other measured variables; for example, adding the corresponding population size (if known) in order to normalize the count later.

\begin{table}[!h]

\caption{\label{tab:tb-sub}A small subset of estimates of tuberculosis burden generated by World Health Organization in 2011 and 2012, with 12 observations and 5 variables. The index refers to column \texttt{year}, the key to multiple columns: \texttt{country} and \texttt{gender}, and the measured variable to column \texttt{count}.}
\centering
\begin{tabular}{lllrr}
\toprule
country & continent & gender & year & count\\
\midrule
Australia & Oceania & Female & 2011 & 120\\
Australia & Oceania & Female & 2012 & 125\\
Australia & Oceania & Male & 2011 & 176\\
Australia & Oceania & Male & 2012 & 161\\
New Zealand & Oceania & Female & 2011 & 36\\
New Zealand & Oceania & Female & 2012 & 23\\
New Zealand & Oceania & Male & 2011 & 47\\
New Zealand & Oceania & Male & 2012 & 42\\
United States of America & Americas & Female & 2011 & 1170\\
United States of America & Americas & Female & 2012 & 1158\\
United States of America & Americas & Male & 2011 & 2489\\
United States of America & Americas & Male & 2012 & 2380\\
\bottomrule
\end{tabular}
\end{table}

The new data structure, tsibble, bridges the gap between raw temporal data and model inputs. Contextual semantics are introduced to tidy data in order to support more intuitive time-related manipulations and enlighten new perspectives for time series model inputs. Index, key and time interval are the three stone pillars to this new semantically structured temporal data. Each is now described in more detail.

\hypertarget{sec:index}{%
\subsection{Index}\label{sec:index}}

Time provides a contextual basis for temporal data. A variable representing time is essential for a tsibble, and is referred to as an ``index''. The ``index'' is an intact data column rather than a masked attribute, which makes time visible and accessible to users. This is highly advantageous when manipulating time. For example, one could easily extract time components, such as time of day and day of week, from the index to visualize seasonal effects of response variables. One could also join other data sources to the tsibble based on common time indexes. The accessibility of the tsibble index motivates data analysis towards transparency and human readability. When the ``index'' is available only as meta information (such as in the \texttt{ts} class), it creates an obstacle for analysts to write these simple queries in a programmatic manner, which should be discouraged from an analytic point of view.

A variable number of time representations can be spotted in the wild. A date-time object, universally accepted across computing systems, is the most commonly used type for representing time. Date-time also typically associates with a time zone including adjustments such as summer time. This diversity and time zone is acknowledged and accommodated by tsibble's index. When creating a tsibble, time indices are arranged from past to future within each series for the strict temporal ordering that is assumed by time series operations.

\hypertarget{sec:key}{%
\subsection{Key}\label{sec:key}}

The ``key'' specification is the second essential ingredient for a tsibble. The ``key'' uniquely identifies observations that are recorded over time in a data table. It is similar to a primary key \autocite{codd_relational_1970} defining each observation in a relational database. In the wide format in which multiple time series are often structured, the columns hold a series of values, so that the column implicitly serves as identification. In long format, columns are melted with names converted to ``key'' values. However, the ``key'' provides much more flexibility. It is not constrained to a single field, but can be composed from multiple fields. The identifying variables from which the ``key'' is constituted remain the same as in the original table with no further tweaks.

The ``key'' is usually known a priori by analysts. For example, \autoref{tab:tb-sub} describes the number of tuberculosis cases for each gender across the countries every year. This data description suggests that columns \texttt{gender} and \texttt{country} have to be declared as the key, similar to a panel variable for longitudinal data. Lacking either of the two will be inadequate, because the observations would not be uniquely identified, and thus a tsibble construction would fail. An alternative specification of the key for this data is to include a third variable \texttt{continent}. Since \texttt{country} is nested within \texttt{continent}, it is a free variable for use. This variable brings additional information that can be used for forecasting reconciliation \autocite{fpp}. The key needs to be explicit when multiple units exist in the data. The key can be implicit when it finds a univariate series in the table, but it cannot be absent from a tsibble.

The ``key'' also provides a link between the data, models, and forecasts. This neatly decouples the data from models and forecasts, leaving more room for necessary model components, such as coefficients, fitted values and residuals. More details are given in \autoref{sec:ts-models}.

\hypertarget{sec:interval}{%
\subsection{Interval}\label{sec:interval}}

One of the cornerstones of time series data, and hence beneath a tsibble, is the time interval. This information plays a critical role in computing statistics (e.g.~seasonal unit root tests) and building models (e.g.~seasonal ARIMA). The principal divide is between regularly or irregularly spaced observations in time. A tsibble permits implicit missing time, making it difficult to distinguish regularity from the index. It relies on a user's specification by switching the \texttt{regular} argument off, when the data involves irregular intervals. This type of data can flow into event-based data modeling, but would need to be processed or regularized to fit models that expect time series.

For data indexed in regular time space, the time interval is automatically calculated, by first computing absolute differences of time indexes and then finding the greatest common divisor. This covers all conceivable cases, assuming that all observations in a tsibble have only one interval. Data collected at different intervals should be organized in separate tsibbles, encouraging well-tailored analysis and models, because each observation may have different underlying data generating processes.

\hypertarget{sec:pipeline}{%
\section{Data pipelines}\label{sec:pipeline}}

A data pipeline describes the flow of data through an analysis, and can generally assist in conceptualizing the process, when it is applied to a variety of problems. \textcite{unix} coined the term ``pipelines'' in software development while developing Unix at Bell Labs. In Unix-based computer operating systems, a pipeline chains together a series of operations on the basis of their standard streams, so that the output of each program becomes the input to another. The Extract, Transform, and Load (ETL) process from recent data warehousing literature dating back to \textcite{kimball2011data} outlines the workflow to prepare data for analysis, and can also be considered a data pipeline. \textcite{viewing-pipeline} describes a viewing pipeline for interactive statistical graphics, that takes control of the transformation from data to plot. \textcite{xgobi}, \textcite{ggobi}, \textcite{orca}, \textcite{plumbing} and \textcite{xie2014reactive} implemented data pipelines for the interactive statistical software \textbf{XGobi}, \textbf{GGobi}, \textbf{Orca}, \textbf{plumbr} and \textbf{cranvas}, respectively. The pipeline is typically described with a one way flow, from data to plot. For interactive graphics, where all plots need to be updated when a user interacts with one plot, the events typically trigger the data pipeline to be run. \textcite{xie2014reactive} uses a reactive programming framework, to implement the pipeline, in which user's interactions trigger a sequence of modules to update their views, that is, practically the same as running the data pipeline producing each plot.

Building a data pipeline is technically difficult: many implementation decisions have to be made about the interface, input and output objects and functionality. The tidy data abstraction lays the plumbing for data analysis modules of transformation, visualization and modeling. Each module communicates with the others, requiring tidy input, producing tidy output, chaining a series of operations together to accomplish the analytic tasks.

What is notable about an effective implementation of a data pipeline is that it coordinates a user's analysis making it cleaner to follow, and permits a wider audience to focus on the data analysis without getting lost in a jungle of computational intricacies. A fluent pipeline glues tidy data and the grammar of data manipulation together. It helps (1) break up a big problem to into manageable blocks, (2) generate human readable analysis workflow, (3) avoid introducing mistakes, at least making it possible to trace them through the pipeline. New data tools developed in the R package \textbf{tsibble} \autocite{R-tsibble} articulate the time series data pipeline, which shepherds raw temporal data through to time series analysis, and plots. More detailed explanations are given in the following sections, and the examples.

\hypertarget{sec:timepipeline}{%
\subsection{Time series transformation}\label{sec:timepipeline}}

\begin{figure}

{\centering \includegraphics[width=\textwidth]{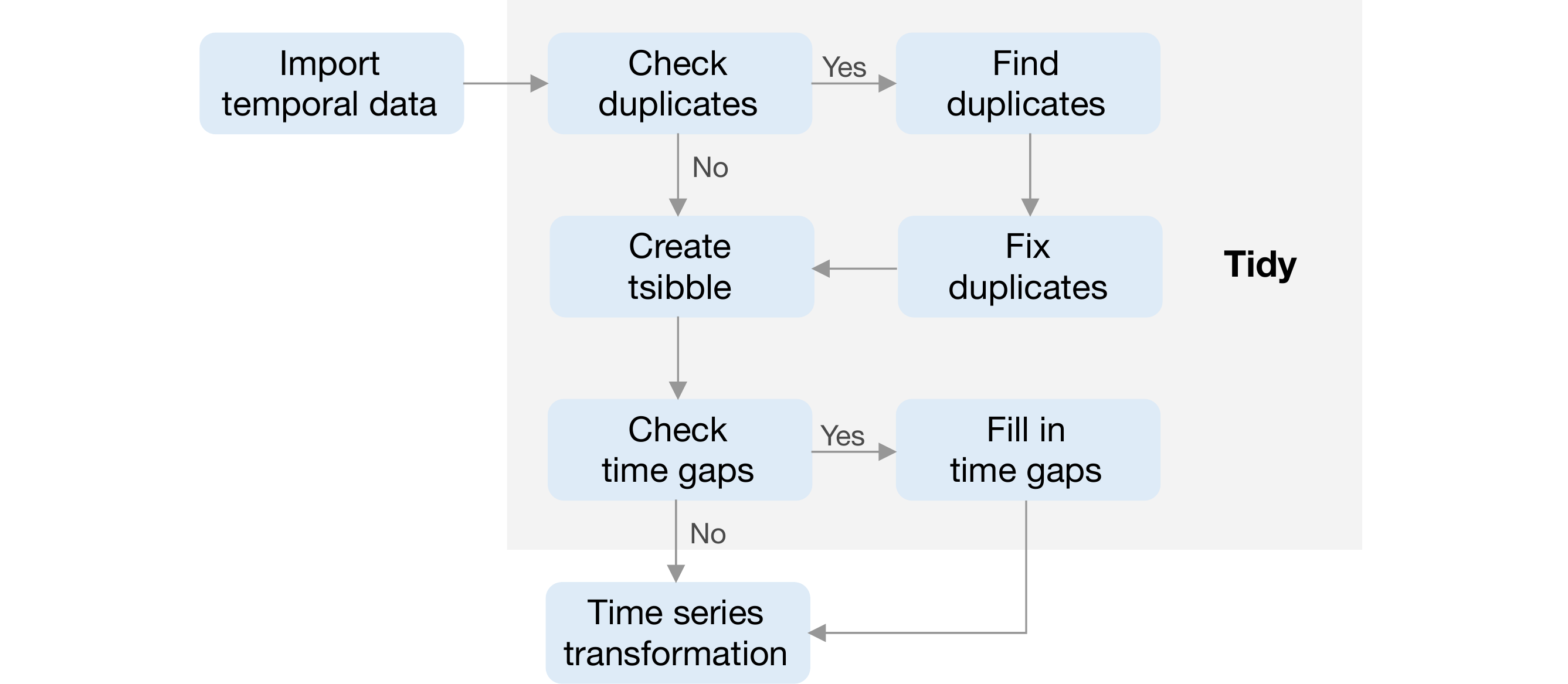} 

}

\caption{A time series pipeline is different from a regular data pipeline: (1) check if the key and index uniquely identify each observation in the data whist creating a tsibble; (2) check if there are time gaps in the tsibble before analysis.}\label{fig:tsibble-pipeline}
\end{figure}

\autoref{fig:tsibble-pipeline} illustrates the distinction of a time series pipeline from a regular data pipeline. It is highly recommended to check for identical entries of key and index before constructing a tsibble. Duplicates signal the data quality issue, which would likely affect subsequent analyses and hence decision making. Analysts are encouraged to gaze at data early and reason about the process of data cleaning. When the data meets the tsibble standard, it flows neatly into the analysis stage and takes full advantage of the tsibble infrastructure.

Many time operations such as lag/lead and time series models, assume an intact vector input ordered in time. Since a tsibble permits time gaps in the index, it is good practice to check and inspect any gaps in time following the creation of a tsibble, in order to prevent inviting these avoidable errors into the analysis. The first suite of verbs (rephrasing actions performed on the object) are provided to understand and tackle implicit missing values: (1) \texttt{has\_gaps()} checks if there exists time gaps; (2) \texttt{scan\_gaps()} reveals all implicit missing observations; (3) \texttt{count\_gaps()} summarizes the time ranges that are absent from the data; (4) \texttt{fill\_gaps()} turns them into explicit ones, along with imputing by values or functions. To look into gaps over individual time periods or full-length time span, the common argument \texttt{.full} in these functions gives an option to easily switch between. The specification of \texttt{.full\ =\ TRUE} will result in fully balanced panels in other words.

\begin{table}[!ht]
\caption{A list of table verbs working with tsibble. Functions in bold originating from the tidyverse and are adapted for tsibble.}\label{tab:verb-table}
\centering
\begin{tabular}{ lll }
\toprule
 & Verb & Description \\ \midrule
\multirow{4}{*}{Time gaps} & \texttt{has\_gaps()} & Test if a tsibble has gaps in time \\
 & \texttt{scan\_gaps()} & Reveal implicit missing entries \\
 & \texttt{count\_gaps()} & Summarize time gaps \\
 & \texttt{fill\_gaps()} & Fill in gaps by values and functions \\ \hline
\multirow{4}{*}{Row-wise} & \textbf{\texttt{filter()}} & Pick rows based on conditions \\
 & \texttt{filter\_index()} & Provide a shorthand for time subsetting \\
 & \textbf{\texttt{slice()}} & Select rows based on row positions \\
 & \textbf{\texttt{arrange()}} & Sort the ordering of row by variables \\ \hline
\multirow{4}{*}{Column-wise} & \textbf{\texttt{select()}} & Pick columns by variables \\
 & \textbf{\texttt{mutate()}} & Add new variables \\
 & \textbf{\texttt{transmute()}} & Drops existing variables \\
 & \textbf{\texttt{summarize()}} & Aggregate values over time \\ \hline
\multirow{3}{*}{Group-wise} & \texttt{index\_by()} & Group by index candidate \\
 & \textbf{\texttt{group\_by()}} & Group by one or more variables \\
 & \texttt{group\_by\_key()} & Group by key variables \\ \hline
\multirow{4}{*}{Reshape} & \textbf{\texttt{gather()}} & Gather columns into long form \\
 & \textbf{\texttt{spread()}} & Spread columns into wide form \\
 & \textbf{\texttt{nest()}} & Nest values in a list-variable \\
 & \textbf{\texttt{unnest()}} & Unnest a list-variable \\ \hline
\multirow{6}{*}{Join tables} & \textbf{\texttt{left\_join()}} & \multirow{6}{*}{Join two tables together} \\
 & \textbf{\texttt{right\_join()}} & \\
 & \textbf{\texttt{full\_join()}} & \\
 & \textbf{\texttt{inner\_join()}} & \\
 & \textbf{\texttt{semi\_join()}} & \\
 & \textbf{\texttt{anti\_join()}} & \\
\bottomrule
\end{tabular}
\end{table}

Besides the time gap verbs, the \textbf{tidyverse} vocabulary is adapted and expanded to facilitate time series transformations, as listed in \autoref{tab:verb-table}. The \textbf{tidyverse} suite showcases the general-purpose verbs for effectively manipulating tabular data, for example \texttt{filter()} picks observations, \texttt{select()} picks variables, and \texttt{left\_join()} joins two tables. But these verbs need handling with care when used in the time series domain. A perceivable difference is summarizing variables between data frame and tsibble using \texttt{summarize()}. The former will reduce to a single summary, whereas the latter will obtain the index and their corresponding summaries. Users who are already familiar with the \textbf{tidyverse}, will experience a gentle learning curve for mastering these verbs and glide into time series analysis with low cognitive load.

Attention has been paid to warning and error handling. The principle that underpins most verbs is a tsibble in and a tsibble out, thereby striving to maintain a valid tsibble by automatically updating index and key under the hood. If the desired temporal ordering is changed by row-wise verbs, a warning is broadcast. If a tsibble cannot be maintained in the output of a pipeline module (likely occurring to column-wise verbs), for example the index is removed by selection, an error informs users of the problem and suggests alternatives. This avoids surprising users and reminds them of the time context.

The tsibble structure and operations support data pipelines for sequencing analysis. \textcite{eopl} asserted ``No matter how complex and polished the individual operations are, it is often the quality of the glue that most directly determines the power of the system.'' Each verb works with other transformation family members in harmony. This set of verbs can result in many combinations to prepare tsibble for a broad range of visualization and modeling problems. Chaining operations is achieved with the pipe operator \texttt{\%\textgreater{}\%} introduced in the \textbf{magrittr} package \autocite{R-magrittr}, read as ``then''. A sequence of functions can be composed in a way that can be naturally read from left to right, which improves the readability of the code. It consequently generates a block of code without saving intermediate values.

Most importantly, a new ecosystem for tidy time series analysis has been undertaken, using the tsibble framework, and is called ``tidyverts'', a play on tidyverse that acknowledges the time series analysis purpose.

\hypertarget{sec:ts-vis}{%
\subsection{Time series visualization}\label{sec:ts-vis}}

As a tsibble is a subclassing of data frame, it integrates well with the grammar of graphics. It is easy to create and extend specialist time series plotting methods based on the tsibble structure, for example autocorrelation plots and calendar-based graphics \autocite{wang-calendarviz-2018}.

\hypertarget{sec:ts-models}{%
\subsection{Time series models}\label{sec:ts-models}}

Modeling is crucial to explanatory and predictive analytics, but often imposes stricter assumptions on tsibble data. The verbs listed in \autoref{tab:verb-table} ease the transition to a tsibble that suits modeling. A tidy forecasting framework built on top of tsibble is under development, which aims at promoting transparent forecasting practices and concise model representation. A tsibble usually contains multiple time series. Batch forecasting will be enabled if a univariate model, such as ARIMA and Exponential Smoothing, is applied to each time series independently. This yields a ``mable'' (short for model table), where each model relates to each ``key'' value in tsibble. This avoids expensive data copying and reduces model storage. The mable is further supplied to forecasting methods, to produce a ``fable'' (short for forecasting table) in which each ``key'' along with its future time holds predictions. It also underlines the advantage of tsibble's ``key'' in acting as linkage between data inputs, models and forecasts. Advanced forecasting techniques, such as vector autocorrelation, hierarchical reconciliation, and ensembles, can be developed in a similar spirit. The modeling module is a current endeavor.

\hypertarget{sec:software}{%
\section{Software structure and design decisions}\label{sec:software}}

\hypertarget{data-first}{%
\subsection{Data first}\label{data-first}}

The primary force that drives the software's design choices is ``data''. All functions in the package \textbf{tsibble} start with \texttt{data} or its variants as the first argument, namely ``data first''. This lays out a consistent interface and addresses the significance of the data throughout the software.

Beyond the tools, the print display provides a quick and comprehensive glimpse of data in temporal context, particularly useful when handling a large collection of data. The contextual summary provided by the print function, shown below on the data from \autoref{tab:tb-sub}, contains (1) data dimension with its shorthand time interval, alongside time zone if date-times, (2) variables that constitute the ``key'' with the number of series. These details aid users in understanding their data better.

\begin{verbatim}
#> # A tsibble: 12 x 5 [1Y]
#> # Key:       country, gender [6]
#>   country     continent gender  year count
#>   <chr>       <chr>     <chr>  <dbl> <dbl>
#> 1 Australia   Oceania   Female  2011   120
#> 2 Australia   Oceania   Female  2012   125
#> 3 Australia   Oceania   Male    2011   176
#> 4 Australia   Oceania   Male    2012   161
#> 5 New Zealand Oceania   Female  2011    36
#> # ... with 7 more rows
\end{verbatim}

\hypertarget{functional-programming}{%
\subsection{Functional programming}\label{functional-programming}}

Rolling window calculations are widely used techniques in time series analysis, and often apply to other applications. These operations are dependent on having an ordering, particularly time ordering for temporal data. Three common types of variations for sliding window operations are:

\begin{enumerate}
\def\labelenumi{\arabic{enumi}.}
\tightlist
\item
  \textbf{slide}: sliding window with overlapping observations.
\item
  \textbf{tile}: tiling window without overlapping observations.
\item
  \textbf{stretch}: fixing an initial window and expanding to include more observations.
\end{enumerate}

\autoref{fig:animate} shows the animations of rolling windows for sliding, tiling and stretching, respectively, on annual tuberculosis cases for Australia. A block of consecutive elements with a window size of 5 are initialized and started rolling sequentially till the end of series by computing average counts.

\begin{figure}

{\centering \animategraphics[width=\textwidth,autoplay,controls,loop]{2}{figure/animate-}{1}{13}

}

\caption[An illustration of window of size 5 computing rolling averages over annual tuberculosis cases in Australia with respect to sliding, tiling and stretching. (Animation needs to be viewed with Adobe Acrobat Reader.)]{An illustration of window of size 5 computing rolling averages over annual tuberculosis cases in Australia with respect to sliding, tiling and stretching. (Animation needs to be viewed with Adobe Acrobat Reader.)}\label{fig:animate}
\end{figure}

Rolling window uses a programming paradigm---functional programming, which is different from those table verbs listed in \autoref{tab:verb-table}. Table verbs expect and return a tsibble, and does what the function name suggests. On the contrary, these rolling window functions could accept arbitrary input types and would return arbitrary sorts of output, depending on which method is put into the rolling window. For example, computing moving averages requires numerics and a function like \texttt{mean()}, and produces averaged numerics. However, rolling window regression takes a data frame and a linear regression method like \texttt{lm()}, and generates a complex object that contains coefficients, fitted values, and etc.

The \textbf{purrr} package \autocite{R-purrr} provides a good example of functional programming in R. It provides a complete and consistent set of tools to iterate each element of a vector with a function. Rolling window does not just iterate but rolls over a sequence of elements, namely \texttt{slide()}, \texttt{tile()} and \texttt{stretch()}. \texttt{slide()} expects one input, \texttt{slide2()} two inputs, and \texttt{pslide()} multiple inputs. For type stability, the functions always return lists. Other variants including \texttt{*\_lgl()}, \texttt{*\_int()}, \texttt{*\_dbl()}, \texttt{*\_chr()} return vectors of the corresponding type, as well as \texttt{*\_dfr()} and \texttt{*\_dfc()} for row-binding and column-binding data frames respectively. Their multiprocessing equivalents prefixed by \texttt{future\_*()} enable rolling in parallel \autocites{R-future}{R-furrr}. This family of functions empowers users to incorporate window-related operations in their workflows.

\hypertarget{modularity}{%
\subsection{Modularity}\label{modularity}}

Modular programming is adopted in the design of the \textbf{tsibble} package. Modularity benefits users by providing small focused and manageable chunks, and provides developers with simpler maintenance.

All user-facing functions can be roughly organized into three major chunks according to their functionality: vector functions (1d), table verbs (2d), and window family. Each chunk is an independent module, but works interdependently. Vector functions in the package mostly deal with time. The atomic functions (such as \texttt{yearmonth()} and \texttt{yearquarter()}) embedded in the \texttt{index\_by()} table verb achieves in collapsing a tsibble to a less granular interval. The substitution of another time function in the \texttt{index\_by()} results in the aggregation of different time resolution. Since these time functions are not exclusive to a tsibble, they can be used in a variety of applications in conjunction with other packages. On the other hand, these tsibble verbs can incorporate many third-party vector functions to step out of the current tsibble zone. It is also generally easier to trace back the errors users encounter from separating 1d and 2d functions, and increase the code readability.

\hypertarget{extensibility}{%
\subsection{Extensibility}\label{extensibility}}

As a fundamental infrastructure, extensibility is a design decision that was employed from the start of \textbf{tsibble}'s development. Contrary to the ``data first'' principle for end users, extensibility is developer focused and would be mostly used in dependent packages, which heavily relies on S3 classes and methods in R \autocite{adv-r}. The package can be extended in two major aspects: custom index and new tsibble class.

Time representation could be arbitrary, for example R's native \texttt{POSIXct} and \texttt{Date} for versatile date-times, nano time for nanosecond resolution implemented in \textbf{nanotime} \autocite{R-nanotime}, and pure numbers in simulations. Ordered factors can also be a source of time, such as month names, January to December, and weekdays, Monday to Sunday. The \textbf{tsibble} package supports an extensive range of index types from numerics to nano time, but there might be custom indexes used for some occasions, for example school semesters. These academic terms vary from one institution to another, within an academic year which is defined differently from a calendar year. A new index would be immediately recognized by the software upon defining \texttt{index\_valid()}, as long as it can be ordered from past to future. The interval regarding semesters is further outlined through \texttt{pull\_interval()}. As a result, the rest of the software methods such as \texttt{has\_gaps()} and \texttt{fill\_gaps()} will have instant support for data that contains this new index.

The class of tsibble is an underlying basis of temporal data, and there is a demand for sub-classing a tsibble. For example, a fable is actually an extension of a tsibble, mentioned in \autoref{sec:ts-models}. A low-level constructor \texttt{new\_tsibble()} provides a vehicle to easily create a new subclass. This new object itself is a tsibble. It perhaps needs more metadata than those of a tsibble, that gives rise to a new data extension, like prediction distributions to a fable. Tsibble verbs are also S3 generics. Developers will be able to implement these verbs for the new class, if necessary.

\hypertarget{tidy-evaluation}{%
\subsection{Tidy evaluation}\label{tidy-evaluation}}

The \textbf{tsibble} packages leverages the \textbf{tidyverse} grammars and pipelines through tidy evaluation \autocite{tidy-eval} via the \textbf{rlang} package \autocite{R-rlang}. In particular, the table verbs extensively use tidy evaluation to evaluate computation in the context of tsibble data and spotlights the ``tidy'' interface that is compatible with the \textbf{tidyverse}. This not only saves a few keystrokes without explicitly repeated references to the data source, but the resulting code is typically cleaner and more expressive.

\hypertarget{sec:cases}{%
\section{Case studies}\label{sec:cases}}

\hypertarget{on-time-performance-for-domestic-flights-in-u.s.a}{%
\subsection{On-time performance for domestic flights in U.S.A}\label{on-time-performance-for-domestic-flights-in-u.s.a}}

The dataset of on-time performance for US domestic flights in 2017 represents event-driven data caught in the wild, sourced from US Bureau of Transportation Statistics \autocite{flights-data}. It contains 5,548,445 operating flights with many measurements (such as departure delay, arrival delay in minutes, and other performance metrics) and detailed flight information (such as origin, destination, plane number and etc.) in a tabular format. This kind of event describes each flight scheduled for departure at a time point in its local time zone. Every single flight should be uniquely identified by the flight number and its scheduled departure time, from a passenger's point of view. In fact, it fails to pass the tsibble hurdle due to duplicates in the original data. An error is immediately raised when attempting to convert this data into a tsibble, and closer inspection has to be carried out to locate the issue. The \textbf{tsibble} package provides tools to easily locate the duplicates in the data with \texttt{duplicates()}. The problematic entries are shown below.

\begin{verbatim}
#>   flight_num  sched_dep_datetime  sched_arr_datetime dep_delay arr_delay
#> 1      NK630 2017-08-03 17:45:00 2017-08-03 21:00:00       140       194
#> 2      NK630 2017-08-03 17:45:00 2017-08-03 21:00:00       140       194
#>   carrier tailnum origin dest air_time distance origin_city_name
#> 1      NK  N601NK    LAX  DEN      107      862      Los Angeles
#> 2      NK  N639NK    ORD  LGA      107      733          Chicago
#>   origin_state dest_city_name dest_state taxi_out taxi_in carrier_delay
#> 1           CA         Denver         CO       69      13             0
#> 2           IL       New York         NY       69      13             0
#>   weather_delay nas_delay security_delay late_aircraft_delay
#> 1             0       194              0                   0
#> 2             0       194              0                   0
\end{verbatim}

The issue was perhaps introduced when updating or entering the data into a system. The same flight is scheduled at exactly the same time, together with the same performance statistics but different flight details. As flight NK630 is usually scheduled at 17:45 from Chicago to New York (discovered by searching the full database), a decision is made to remove the first row from the duplicated entries before proceeding to the tsibble creation.

This dataset is intrinsically heterogeneous, encoded in numbers, strings, and date-times. The tsibble framework, as expected, incorporates this type of data without any loss of data richness and heterogeneity. To declare the flight data as a valid tsibble, column \texttt{sched\_dep\_datetime} is specified as the ``index'', and column \texttt{flight\_num} as the ``key'' via \texttt{id(flight\_num)}. As a result of event timing, the data are irregularly spaced, and hence switching to the irregular option is necessary. The software internally validates if the key and index produce distinct rows, and then sorts the key and the index from past to recent. When the tsibble creation is done, the print display is data-oriented and contextually informative, including dimensions, irregular interval (\texttt{5,548,444\ x\ 22\ {[}!{]}\ \textless{}UTC\textgreater{}}) and the number of time-based observational units (\texttt{flight\_num\ {[}22,562{]}}).

\begin{verbatim}
#> # A tsibble: 5,548,444 x 22 [!] <UTC>
#> # Key:       flight_num [22,562]
\end{verbatim}

Transforming a tsibble for exploratory data analysis with a suite of time-specific and general-purpose manipulation verbs can result in well-constructed pipelines. From the perspective of a passenger, for example, one needs to travel smart by choosing an efficient carrier to fly with and the time of day to avoid congestion. To explore this data, we drill down starting with annual carrier performance and followed by disaggregation to finer time resolutions.

\begin{figure}

{\centering \includegraphics[width=\textwidth]{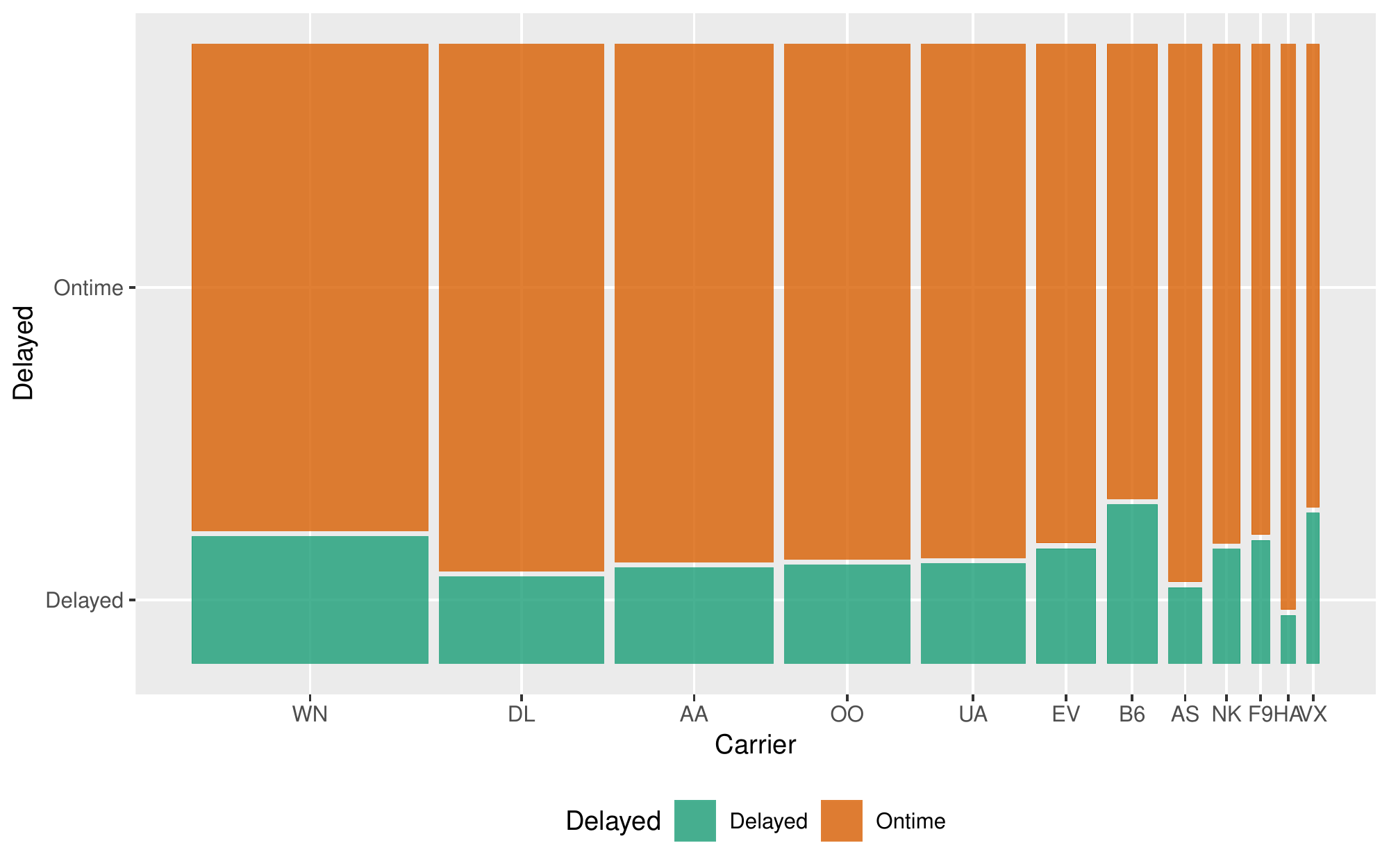} 

}

\caption{Mosaic plot showing the association between the size of airline carriers and the delayed proportion of departures in 2017. Southwest Airlines is the largest operator, but does not operate as efficiently as Delta. Hawaiian Airlines, a small operator, outperforms the rest.}\label{fig:carrier-mosaic}
\end{figure}

\begin{figure}

{\centering \includegraphics[width=\textwidth]{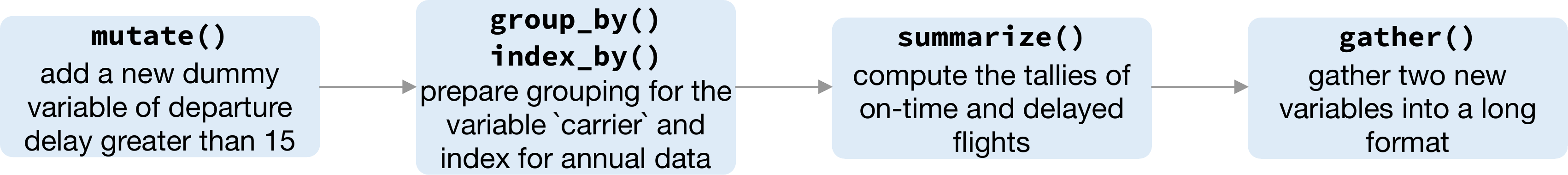} 

}

\caption{Flow chart illustrating the pipeline that preprocessed the data for creating \autoref{fig:carrier-mosaic}.}\label{fig:flight-pipeline1}
\end{figure}

\autoref{fig:carrier-mosaic} visually presents the end product of aggregating the number of on-time and delayed flights to the year interval by carriers. This pipeline is initialized by defining a new variable if the flight is delayed, and involves summarizing the tallies of on-time and delayed flights for each carrier annually. To prepare the summarized data for a mosaic plot, it is further manipulated by melting new tallies into a single column. The flow chart (\autoref{fig:flight-pipeline1}) demonstrates the operations undertaking in the data pipeline. The input to this pipeline is a tsibble of irregular interval, and the output ends up with a tsibble of unknown interval (as each carrier ends up with only one annual summary). The final data set includes each carrier along with a single year, with the interval undetermined, which in turn feeds into the mosaic display. Note that Southwest Airlines (WN), as the largest carrier, operates less efficiently than Delta (DL).

\begin{figure}

{\centering \includegraphics[width=\textwidth]{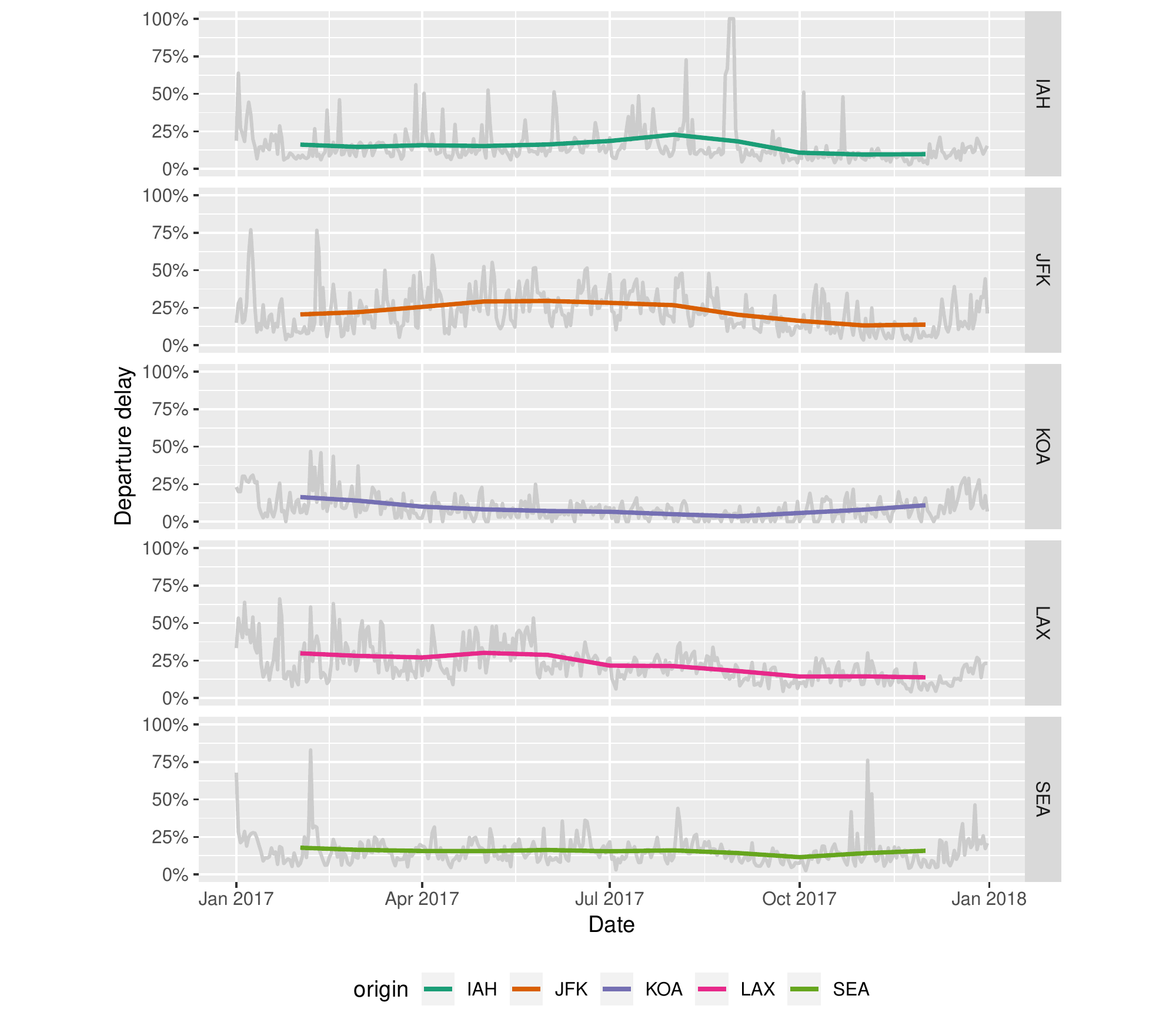} 

}

\caption{Daily delayed percentages for departure with two-month moving averages overlaid at five international airports. There are least fluctuations and relatively fewer delays observed at KOA airport. The estimates of temporal trend are around 25\% across the other four airports, but highlight different time periods of severe delays.}\label{fig:sel-monthly-plot}
\end{figure}

A closer examination of some big airports across the US will give an indication of how well the busiest airports manage the outflow traffic on a daily basis. A subset that contains observations for Houston (IAH), New York (JFK), Kalaoa (KOA), Los Angeles (LAX) and Seattle (SEA) airports is obtained first. The succeeding operations compute delayed percentages every day at each airport, which are framed as grey lines in Figure \ref{fig:sel-monthly-plot}. Winter months tend to fluctuate a lot compared to the summer across all the airports. Superimposed on the plot are two-month moving averages, so the temporal trend is more visible. The number of days for each month is variable. Moving averages for two months call for computing weighted mean. But this can also be accomplished using a pair of commonly used verbs--\texttt{nest()} and \texttt{unnest()} to handle list-columns, without weight specification. The sliding operation with a large window size smooths out the fluctuations and gives a stable trend around 25\% over the year. LAX airport has seen a gradual decline in delays over the year, whereas the SEA airport has a steady number delays over time. The IAH and JFK airports have more delays in the middle of year, while the KOA has the inverse pattern with higher delay percentage in both ends of the year.

\begin{figure}

{\centering \includegraphics[width=\textwidth]{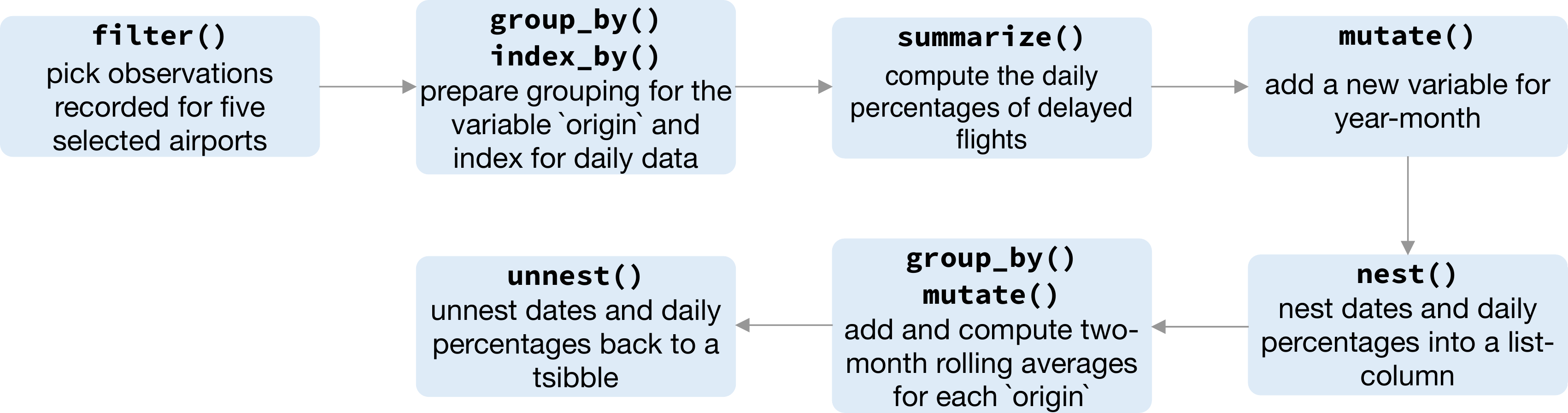} 

}

\caption{Flow chart illustrating the pipeline that preprocessed the data for creating \autoref{fig:sel-monthly-plot}.}\label{fig:flight-pipeline2}
\end{figure}

\begin{figure}

{\centering \includegraphics[width=\textwidth]{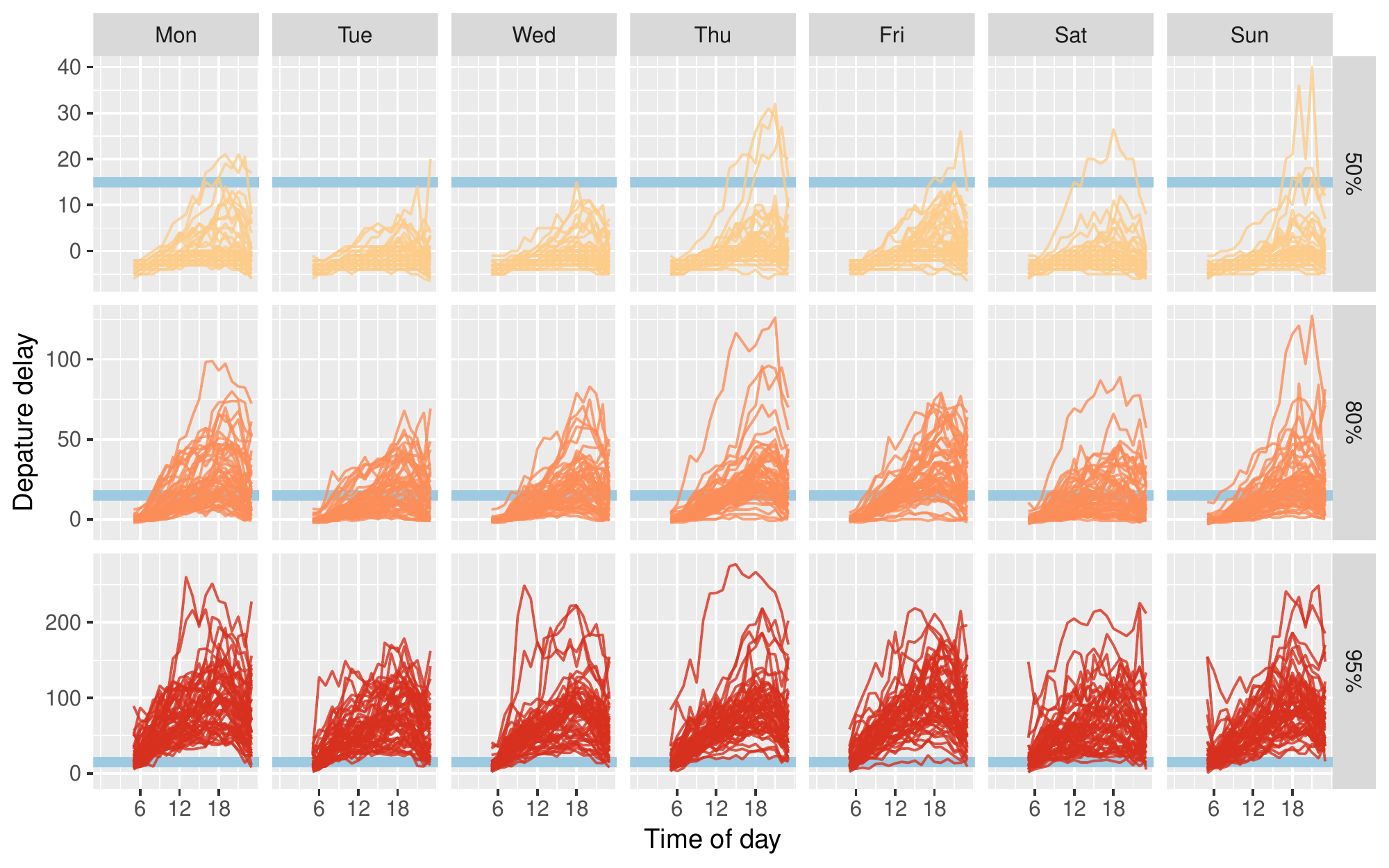} 

}

\caption{Small multiples of lines about departure delay against time of day, faceting day of week and 50\%, 80\% and 95\% quantiles. A blue horizontal line indicates the 15-minute on-time standard to help grasp the delay severity. Passengers are apt to hold up around 18 during a day, and are recommended to travel early. The variations increase substantially as the upper tails.}\label{fig:draw-qtl}
\end{figure}

What time of day and day of week should we travel to avoid suffering from horrible delay? \autoref{fig:draw-qtl} plots hourly quantile estimates across day of week in the form of small multiples. The upper-tail delay behaviors are of primary interest, and hence 50\%, 80\% and 95\% quantiles are shown. To reduce the likelihood of suffering a delay, it is recommended to avoid the peak hour around 6pm (18).

\begin{figure}

{\centering \includegraphics[width=\textwidth]{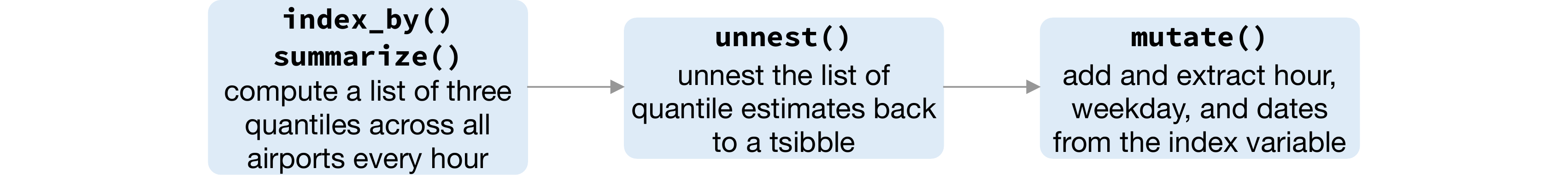} 

}

\caption{Flow chart illustrates the pipeline that preprocesses the data for creating \autoref{fig:draw-qtl}.}\label{fig:flight-pipeline3}
\end{figure}

\FloatBarrier

\hypertarget{smart-grid-customer-data-in-australia}{%
\subsection{Smart-grid customer data in Australia}\label{smart-grid-customer-data-in-australia}}

Sensors have been installed in households across major cities in Australia to collect data for the smart city project. One of the trials is monitoring households' electricity usage through installed smart meters in the area of Newcastle over 2010--2014 \autocite{smart-meter}. Data from 2013 have been sliced to examine temporal patterns of customer's energy consumption with \textbf{tsibble} for this case study. Half-hourly general supply in kwH have been recorded for 2,924 customers in the data set, resulting in 46,102,229 observations in total. Daily high and low temperatures in Newcastle in 2013 provides explanatory variables other than time in a different data table \autocite{newcastle-temp}, obtained using the R package \textbf{bomrang} \autocite{R-bomrang}. Two data tables might be joined to explore how local weather can contribute to the variations of daily electricity use when needed.

\begin{figure}

{\centering \includegraphics[width=\textwidth]{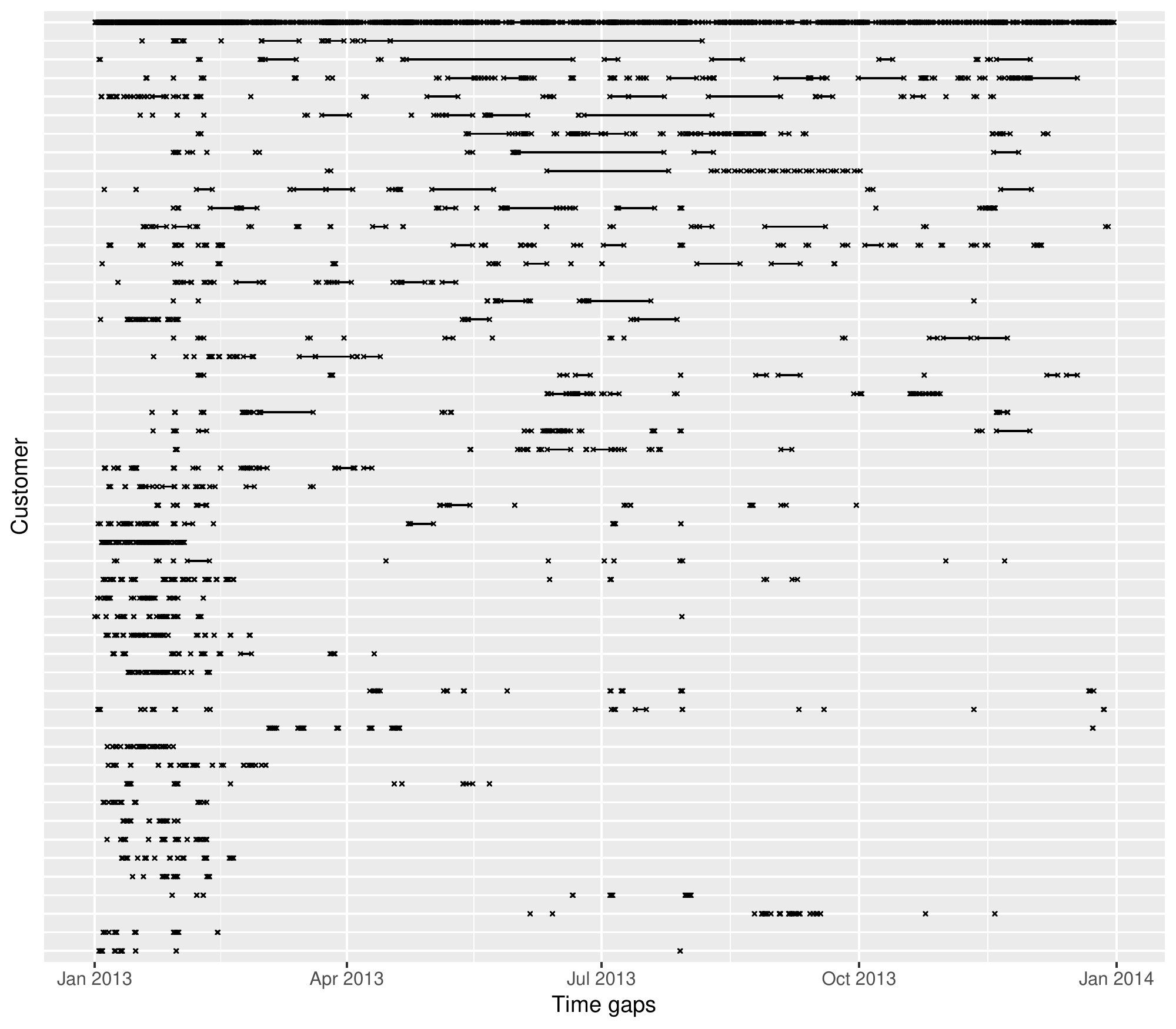} 

}

\caption{Time gap plots for the 49 customers with most implicit missing values, and the remaining customers grouped into the one line at top. Each cross represents an observation missing in time and a line between two dots shows continuous missingness over time. Each row corresponds to one customer. Missing values occur at various times, with more in January and February than other months.}\label{fig:count-gaps}
\end{figure}

During a power outage, electricity usage for some households may become unavailable, thus resulting in implicit missing values in the database. Gaps in time occur to 17.9\% of the households in this dataset. It would be interesting to explore these missing patterns as part of a preliminary analysis. Since the smart meters have been installed at different dates for each household, it is reasonable to assume that the records are obtainable for different time lengths for each household. \autoref{fig:count-gaps} shows the gaps for the top 49 households arranged in rows from high to low in tallies. (The remaining households values have been aggregated into a single batch and appear at the top.) Missing values can be seen to occur at any time during the entire span. A small number of customers have undergone energy unavailability in consecutive hours, indicated by a line range in the plot. On the other hand, the majority suffer occasional outages with more frequent occurrence in January.

\begin{figure}

{\centering \includegraphics[width=\textwidth]{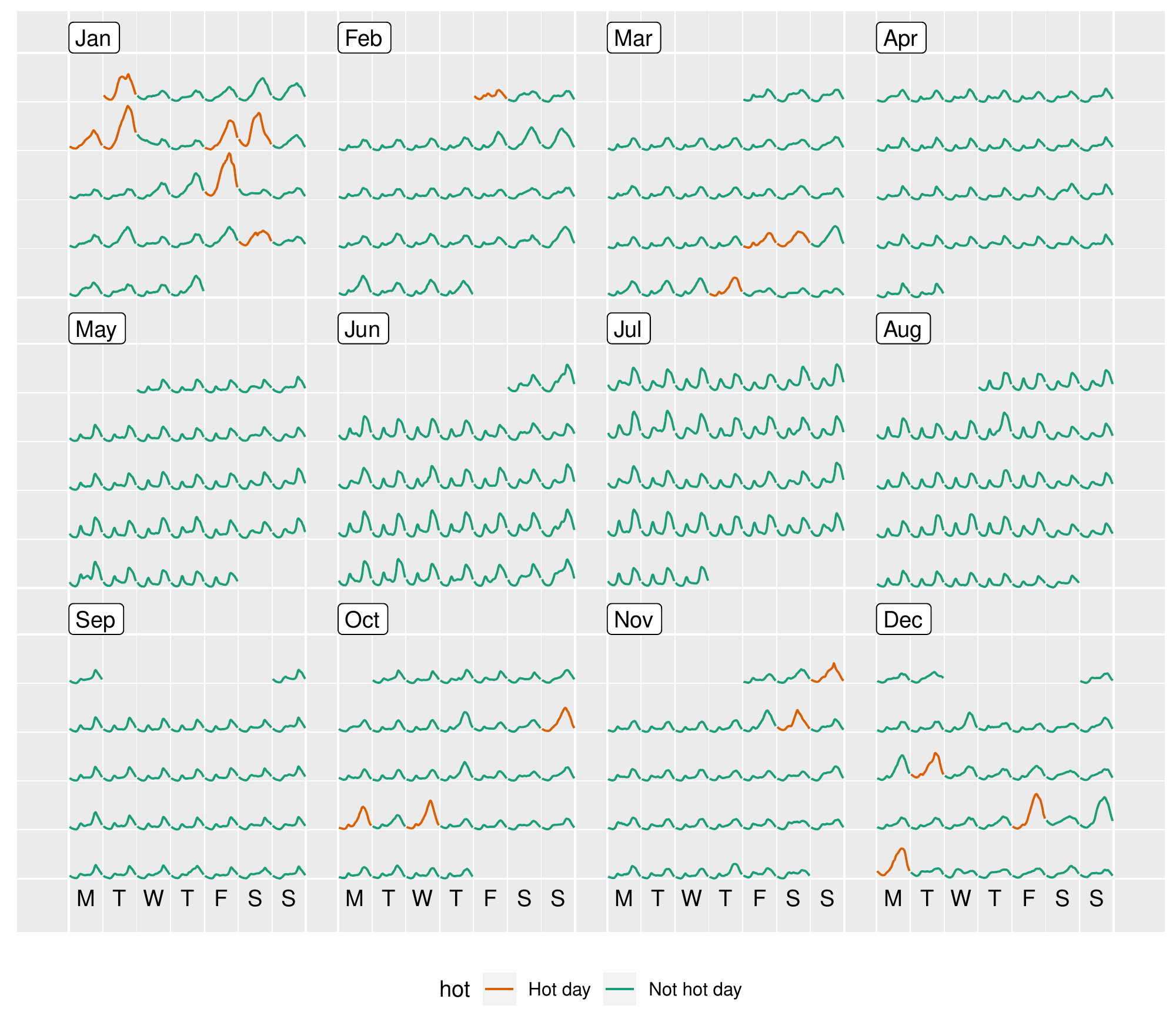} 

}

\caption{Half-hourly average electricity use across all customers in the region, organized into calendar format, with color indicating hot days. Energy use of hot days tends to be higher, suggesting air conditioner use. Days in the winter months have a double peak suggesting heater use.}\label{fig:calendar-plot}
\end{figure}

Aggregation across all individuals helps to sketch a big picture of the behavioral change over time, organized into a calendar display (\autoref{fig:calendar-plot}). Each glyph represents the daily pattern of average residential electricity usage every thirty minutes. Higher consumption is indicated by higher values, and typically occurs in daylight hours. Color indicates hot days. The daily snapshots vary depending on the season in the year. During the summer months (December and January), the late-afternoon peak becomes predominant driven by the use of air conditioning, especially on hot days with daily average temperature greater than 25 degrees C. However, the winter time (July and August) sees two peaks in a day, which is probably due to heating in the morning and evening. This plot illustrates how the tsibble data can easily integrate with other tools and graphics.

\begin{figure}

{\centering \includegraphics[width=\textwidth]{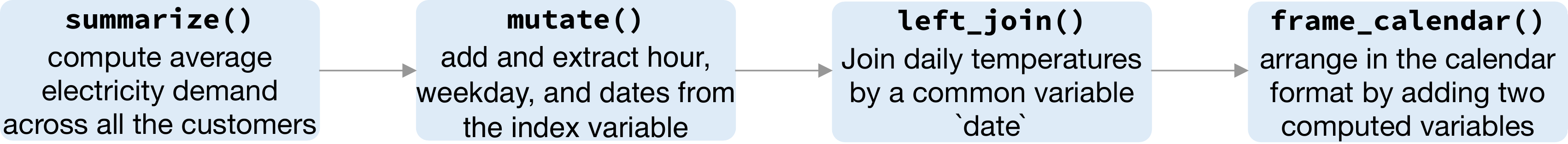} 

}

\caption{Flow chart illustrating the pipeline that preprocessed the data for creating \autoref{fig:calendar-plot}.}\label{fig:elec-pipeline2}
\end{figure}

\hypertarget{sec:conclusion}{%
\section{Conclusion and future work}\label{sec:conclusion}}

The ``tsibble'' is a new data abstraction to represent temporal data, allowing the ``tidy data'' principles to be brought to the time domain. Tidy data begins to take shape in the state of time with the introduction of the contextual semantics of index and key. A declared index provides direct support to the time variable; variables that comprise the key define observations over time. These semantics further determine unique data entries required for a valid tsibble. No matter how temporal data arrives, a tsibble respects a time index and maintains the data richness. A tsibble frictionlessly allows transformation, visualization and modeling, and smoothly shifts between them, allowing for rapid iteration to gain data insights.

A missing piece of the \textbf{tsibble} package is to enable user-defined calendars and to respect structurally missing observations. For example, a call center may operate only between 9:00 am and 5:00 pm on week days, and stock trading resumes on Monday straight after Friday. No data available outside trading hours would be labeled as structural missingness, which tsibble currently disregards. However, a few R packages provide functionality to create and manage many sorts of calendars, including market-specific business calendars. Generally, custom calendars are easily embedded into the tsibble framework. Consequently these tsibble operators, like \texttt{fill\_gaps()}, would work out of the box, and forecasts would be generated within its definable time range.

The \textbf{tsibble} package provides the grammar of temporal data manipulation, regardless of how the data is stored. Currently, it works for managing and manipulating temporal data frames in memory locally. But it is possible to work with remote tables stored in databases, such as SQLite and MySQL, using exactly the same tsibble code. This is left for future work.

\hypertarget{acknowledgments}{%
\section*{Acknowledgments}\label{acknowledgments}}
\addcontentsline{toc}{section}{Acknowledgments}

The authors would like to thank Mitchell O'Hara-Wild for many discussions on the software development and Davis Vaughan for contributing ideas on rolling window functions. We also thank Stuart Lee for the feedback on this manuscript. This article was created with knitr \autocite{knitr} and R Markdown \autocite{rmarkdown}. The project's Github repository \url{https://github.com/earowang/paper-tsibble} houses all materials required to reproduce this article and a history of the changes.

\printbibliography

\end{document}